\documentstyle[prd,aps,epsf,amsmath,amsthm,amssymb]{revtex}

\newcommand{\half}{\mbox{$\textstyle \frac{1}{2}$}}
\newcommand{\quat}{\mbox{$\textstyle \frac{1}{4}$}}

\begin{document} 

\tighten
\draft
\preprint{DAMTP-1999-??}
\twocolumn[\hsize\textwidth\columnwidth\hsize\csname 
@twocolumnfalse\endcsname

\title{Stochastic Reduction in Nonlinear Quantum Mechanics} 

\author{
Dorje C. Brody$^{*}$ and Lane P. Hughston$^{\dagger}$
} 

\address{* The Blackett Laboratory, Imperial College, 
London SW7 2BZ, UK} 

\address{$\dagger$ Department of Mathematics, King's College 
London, The Strand, London WC2R 2LS, UK} 

\date{\today} 

\maketitle 

\begin{abstract}
Stochastic extensions of the Schr\"odinger equation have attracted 
attention recently as plausible models for state reduction in 
quantum mechanics. Here we formulate a general approach to 
stochastic Schr\"odinger dynamics in the case of a nonlinear state 
space of the type proposed by Kibble. We derive a number of new 
identities for observables in the nonlinear theory, and establish 
general criteria on the curvature of the state space sufficient to 
ensure collapse of the wave function. 
\end{abstract} 


\vskip2pc] 

A generalisation of quantum mechanics was considered by Mielnik 
\cite{mielnik}, who introduced the notion of nonlinear observables. 
Two alternative extensions of the standard quantum theory 
were then proposed by Kibble \cite{kibble}. The first alternative 
is based on the phase space formulation of quantum mechanics. 
Here, if we work with the space of rays through the origin of 
Hilbert space, then the Schr\"odinger equation reduces to 
Hamilton's equation of classical mechanics \cite{etc}, 
except that the quantum Hamiltonian is of a special restricted 
form. Thus a natural generalisation is to remove this constraint. 
When such trajectories are lifted from the space of rays to Hilbert 
space, we obtain nonlinear wave equations. 

The general properties of nonlinear observables were subsequently 
analysed in detail by Weinberg \cite{weinberg}. Following this, it 
was pointed out by Gisin \cite{gisin} that the evolution of the 
density matrix is not autonomous in the nonlinear mechanics of 
\cite{kibble,weinberg}, and that this may be physically undesirable. 
However, it was also indicated in \cite{gisin} that there is another 
type of nonlinear quantum dynamics for which the evolution of the 
density matrix 
is autonomous, and a number of desirable features of linear evolution 
are extended in a natural way. This is the stochastic dynamics 
developed by Pearle and others \cite{pearl}. These dynamics are 
of significance because they exhibit natural reductive 
properties: starting from a given initial state, the 
system evolves stochastically in such a way to ensure collapse to 
an eigenstate of one or more designated observables. 

Kibble's second alternative for a nonlinear quantum theory is in 
essence to consider a general K\"ahler manifold as the phase space 
of quantum mechanics, instead of the space of rays. The idea is that, 
in the presence of interactions, the states accessible to a quantum 
system constitute a curved space ${\mathfrak M}$ 
which has the structure of 
a complex manifold endowed with a compatible symplectic structure. 
The dynamics of the state are then governed by a Hamiltonian flow 
which is also an isometry. 

In the present article, we consider stochastic state reduction 
models within the framework of Kibble's second theory. The 
advantage of a stochastic dynamics in this context is that it 
leads to a probabilistic interpretation, a feature hitherto missing 
in the nonlinear theory. Remarkably, many of the key features of 
the basic stochastic reduction models carry through to a fully 
nonlinear state space. 
The main results are to determine general criteria sufficient to 
ensure state reduction in the nonlinear theory, and to express 
these criteria directly in terms of geometrical features of the 
state manifold. Thus it is the geometry of 
the quantum state manifold that determines whether reduction 
takes place, and if so, how rapidly. 

After introducing the relevant state space geometry and elements 
of stochastic calculus on manifolds, a number of identities 
concerning the properties of quantum observables are established 
in Lemmas 1-5. These results are then applied to formulate 
general theorems governing reduction processes on nonlinear state 
spaces.   

Let us first recall briefly the phase space formulation of quantum 
theory. We consider a finite dimensional complex Hilbert 
space ${\cal H}$ of which a typical element is 
denoted $\psi^{\alpha}$ ($\alpha=0, 1, \cdots, n$). Given 
the Hamiltonian operator $H^{\alpha}_{\beta}$, the dynamics of the 
state is determined by the Schr\"odinger equation ${\rm i}\hbar 
\partial_{t}\psi^{\alpha} = H^{\alpha}_{\beta}\psi^{\beta}$. The 
expectation $F^{\alpha}_{\beta}{\bar\psi}_{\alpha}\psi^{\beta}/
{\bar\psi}_{\gamma}\psi^{\gamma}$ of an observable 
$F^{\alpha}_{\beta}$ in the state $\psi^{\alpha}$ is invariant 
under the scale transformations 
$\psi^{\alpha}\rightarrow\lambda\psi^{\alpha}$ 
($\lambda\in{\mathbb C}-\{0\}$). Hence we can work with the space 
of equivalent classes of state vectors modulo such transformations, 
i.e., the complex projective space $P^{n}$. 

We regard $P^{n}$ as a real manifold $\Gamma$ of dimension $2n$. 
It is known that $\Gamma$ has a natural symplectic structure 
$\omega_{ab}$, as well as a Riemannian structure given by the 
Fubini-Study metric $g_{ab}$. These two structures are 
compatible in the sense that there exists an 
integrable complex structure $J^{a}_{\ b}$ on $\Gamma$, satisfying 
$J^{a}_{\ c}J^{c}_{\ b}=-\delta^{a}_{\ b}$, such that 
$\nabla_{a}J^{b}_{\ c}=0$ and $g^{ac}\omega_{cb}=J^{a}_{\ b}$, 
where $\nabla_{a}$ is the covariant derivative associated 
with $g_{ab}$, and $g_{ac}g^{cb}=\delta_{a}^{\ b}$. We use Roman 
indices $(a,b,\cdots)$ for tensorial operations on $\Gamma$. The 
compatibility conditions make $\Gamma$ a K\"ahler manifold. 

The special feature that identifies $\Gamma$ as the quantum phase 
space is that the Schr\"odinger equation can be expressed in the 
Hamiltonian form $\hbar dx^{a}/dt = 2 \omega^{ab}\nabla_{b}H(x)$. 
Here $\omega^{ab}=g^{ac}g^{bd}\omega_{cd}$ and $H(x)$ is the 
Hamiltonian function on $\Gamma$, given by the expectation of the 
operator $H^{\alpha}_{\beta}$ in the equivalence class of state 
vectors corresponding to the point $x\in\Gamma$. 
The vector $Z^{a}=\omega^{ab}\nabla_{b}H$ tangent to the 
Schr\"odinger trajectory satisfies the Killing equation 
$\nabla_{(a}Z_{b)}=0$ iff  $H(x)$ is the expectation of a quantum 
observable in the state $x$. Therefore, the Schr\"odinger 
evolution preserves the distance, and hence the transition 
probability, between any given two states. More generally, any 
isometry of $\Gamma$ is a Hamiltonian flow associated with a 
quantum observable. The energy eigenstates are the fixed points of 
the flow, at which $\nabla_{a}H=0$. The first 
alternative of Kibble is to replace the observable $H(x)$ by a 
general function on $\Gamma$. Then the resulting trajectories are 
Hamiltonian but no longer Killing, and the implied dynamics 
on ${\cal H}$ is governed by a nonlinear wave equation. This is 
not the generalisation we consider here. 

For the consideration of Kibble's second alternative, it will 
be useful first to develop a differential geometric framework for 
the standard operations of quantum mechanics (we set $\hbar=1$). 
If $F(x)$ and $G(x)$ are observables, the expectation of 
their commutator is also an observable, given by the 
Poisson bracket $2\omega^{ab}\nabla_{a}F\nabla_{b}G$. Then if 
$x_{t}$ is a Schr\"odinger trajectory and $F_{t}=F(x_{t})$, it 
follows that $dF_t=2\omega^{ab}\nabla_{a}F\nabla_{b}Hdt$, 
where $H$ is the Hamiltonian. This tells us how the expectation 
of $F$ changes along the flow generated by $H$. For any 
observable $F(x)$, we define the associated dispersion by  
\begin{eqnarray}
V^{F}=g^{ab}\nabla_{a}F\nabla_{b}F .  
\label{eq:2}
\end{eqnarray}
In the linear theory $V^{F}(x)$ is the squared uncertainty of 
$F$ in the state $x$. As a consequence of the inequality 
$(g_{ab}X^{a}X^{b})(g_{ab}Y^{a}Y^{b})\geq(g_{ab}X^{a}Y^{b})^{2}
+(\omega_{ab}X^{a}Y^{b})^{2}$, which holds for all $X^{a}$ and 
$Y^{a}$, we obtain the Heisenberg relation $V^{F}V^{G}\geq
(\omega^{ab}\nabla_{a}F\nabla_{b}G)^{2}$ if we set 
$X^{a}=\omega^{ab}\nabla_{b}F$ and $Y^{a}=\omega^{ab}\nabla_{b}G$. 

Kibble's second alternative for a nonlinear quantum theory is to 
let the state space be a general K\"ahler manifold ${\mathfrak M}$, 
with metric 
$g_{ab}$, symplectic structure $\omega_{ab}$, and complex structure 
$J_{a}^{\ b}$. In the nonlinear theory we say that a real function 
$F(x)$ on ${\mathfrak M}$ is an observable iff the corresponding 
Hamiltonian vector field $X^{a}=\omega^{ab}\nabla_{b}F$ is an 
isometry. This agrees with the usual characterisation of 
observables when ${\mathfrak M}=\Gamma$. We note that if 
${\mathfrak M}$ is compact and has vanishing first Betti number, 
then any Killing field on ${\mathfrak M}$ is Hamiltonian, with at 
least two distinct eigenstates \cite{frankel}. As in the 
linear theory, we interpret $F(x)$ as the expectation of the 
result of a measurement of the given observable in the state 
$x\in{\mathfrak M}$. 

{\bf Lemma 1}. {\sl If $F(x)$ and $G(x)$ are observables, then 
their commutator is an observable.} 

The proof is as follows. If $X^{a}$ and $Y^{a}$ are Hamiltonian 
flows, then their Lie bracket is $X^{b}\nabla_{b}Y^{a}-Y^{b}
\nabla_{b}X^{a}=\omega^{ab}\nabla_{b}(\omega_{cd}X^{c}Y^{d})$. 
If $F(x)$ and $G(x)$ are generators of $X^{a}$ and $Y^{a}$, then 
$\omega_{cd}X^{c}Y^{d}=\omega^{cd}\nabla_{c}F\nabla_{d}G$. 
Furthermore, if $X^{a}$ and $Y^{a}$ are Killing, so is their 
Lie bracket. Therefore, the Hamiltonian flow generated by the 
commutator of $F(x)$ and $G(x)$ is Killing. As a consequence 
we obtain also the following nonlinear generalisation of  
an identity due to Adler and Horwitz \cite{adler}: 

{\bf Lemma 2}. {\sl If $F(x)$ and $G(x)$ are observables, then 
$\nabla_{b}F\nabla^{b}\nabla^{a}G-\nabla_{b}G\nabla^{b}
\nabla^{a}F=\omega^{ab}\nabla_{b}(\omega^{cd}\nabla_{c}F
\nabla_{d}G)$.} 

As in the linear theory, the eigenstates of an observable $F(x)$ 
are the points of ${\mathfrak M}$ at which $\nabla_{a}F=0$. 
The value of $F(x)$ at a critical point 
is the corresponding eigenvalue. In the nonlinear theory 
$V^{F}(x)$ does not in general have an interpretation as a 
moment, but nevertheless remains a measure of the dispersion 
of $F(x)$ in the given state. In particular, the Heisenberg 
relation holds. 

The Schr\"odinger trajectories in the nonlinear theory are generated 
by a Hamiltonian $H(x)$, which we assume to be an observable. The 
following result shows that for any observable commuting with the 
Hamiltonian, its dispersion is constant along the Schr\"odinger 
trajectory. This fact will be used later to derive the stochastic 
dynamics of the energy dispersion. 

{\bf Lemma 3}. {\sl If $F$ is an observable that commutes 
with the Hamiltonian $H$, then 
$\omega^{ab}\nabla_{a}H\nabla_{b}V^{F}=0$. } 

The proof is as follows. Equation (\ref{eq:2}) implies that 
$\omega^{ab}\nabla_{a}H\nabla_{b}V^{F}=2\omega^{ab}\nabla_{a}H
\nabla_{b}\nabla_{c}F\nabla^{c}F$, and thus 
\begin{eqnarray} 
\omega^{ab}\nabla_{a}H\nabla_{b}V^{F} &=& 
2\nabla_{c}(\omega^{ab}\nabla_{a}H\nabla_{b}F)\nabla^{c}F 
\nonumber \\ & & \ -2\nabla_{c}(\omega^{ab}\nabla_{a}H)
\nabla_{b}F\nabla^{c}F . 
\label{eq:3}
\end{eqnarray}
The first term on the right vanishes because $H$ and $F$ commute, 
whereas the second term vanishes on account of the Killing 
equation satisfied by $Z^{a}$. 

To proceed further we now introduce briefly the elements of 
stochastic differential geometry. The basic 
process we consider is the Wiener process $W_{t}$ defined on 
a probability space $({\sl\Omega}, {\cal F},{\bf P})$, where 
${\sl\Omega}$ is the sample space, ${\cal F}$ denotes the 
filtration, and ${\bf P}$ is the probability measure. We say 
that $W_{t}$ is a Wiener process if it is continuous, $W_{0}=0$, 
$W_{t}-W_{s}$ ($0\leq s<t$) is independent of the information 
${\cal F}_{s}$ up to time $s$, and $W_{t}-W_{s}$ is normally 
distributed with mean zero and variance $t-s$. A process 
$\sigma_{t}$ is said to be adapted to the filtration 
${\cal F}_{t}$ generated by $W_{t}$ if its random value at time 
$t$ is determined by the history of $W_{t}$ up to that time. If 
$\sigma_{t}$ is ${\cal F}_{t}$-adapted, then the stochastic 
integral $M_{t} = \int_{0}^{t} \sigma_{s} dW_{s}$ exists, 
provided $\sigma_{t}$ is almost surely square-integrable. 
If the variance of $M_{t}$ exists, then $M_{t}$ satisfies 
the martingale conditions ${\mathbb E}[|M_{t}|]<\infty$ and 
${\mathbb E}[M_{t}|{\cal F}_{s}]=M_{s}$, where ${\mathbb E}[-]$ 
denotes expectation with respect to ${\bf P}$. The second condition 
implies that, given the history of the Wiener process up to time $s$, 
the expectation of $M_t$ for $t\geq s$ is given by its value at $s$. 
The variance of $M_{t}$ is determined by the Ito isometry 
${\mathbb E}[M_{t}^{2}]={\mathbb E}[\int_{0}^{t}\sigma_{s}^{2}ds]$.

A general Ito process is defined by the integral 
$x_{t}=x_{0}+\int_{0}^{t}\mu_{s}ds+\int_{0}^{t}\sigma_{s}dW_{s}$, 
where the adapted processes $\mu_{t}$ and $\sigma_{t}$ are called 
the drift and the volatility of $x_{t}$. A convenient way to express 
this is to write $dx_{t}=\mu_{t}dt+\sigma_{t}dW_{t}$. 
In the special case $\mu_{t}=\mu(x_t)$ and $\sigma_t=
\sigma(x_{t})$, where $\mu(x)$ and $\sigma(x)$ are prescribed 
functions, the process $x_t$ is said to be a diffusion.  

This analysis can be generalised to the case of a diffusion 
$x_{t}$ taking values on a manifold 
${\mathfrak M}$, driven by a standard $m$-dimensional Wiener 
process $W_{t}^{i}$ ($i=1,2,\cdots,m$). Let $\nabla_{a}$ be a 
torsion-free connection on ${\mathfrak M}$, and suppose 
$\mu^{a}(x)$ and $\sigma^{a}_{i}(x)$ are $m+1$ vector fields on 
${\mathfrak M}$. Then for the general diffusion 
on ${\mathfrak M}$ we have the stochastic equation $dx^{a}=
\mu^{a}dt+\sigma^{a}_{i}dW^{i}_{t}$, where $dx^{a}$ is the 
covariant Ito differential. The quadratic relation 
$dx^{a}dx^{b}=h^{ab}dt$, where $h^{ab}=\sigma_{i}^{a}
\sigma^{bi}$, follows from the identities 
$dt^2=0$, $dtdW_{t}^i=0$, and $dW_{t}^{i}dW_{t}^{j}=
\delta^{ij}dt$. If we choose 
$\mu^{a}=2\omega^{ab}\nabla_{b}H$, $\sigma^{a}_{i}=0$, and 
${\mathfrak M}=\Gamma$, then 
$x_{t}$ reduces to the Schr\"odinger evolution. 

For any smooth function $\phi(x)$ on ${\mathfrak M}$ we 
define the process $\phi_{t}=\phi(x_{t})$, and Ito's formula 
takes the form $d\phi_{t}=(\nabla_{a}\phi)dx^{a}+\frac{1}{2}
(\nabla_{a}\nabla_{b}\phi)dx^{a}dx^{b}$, or, more explicitly, 
\begin{eqnarray}
d\phi=\left(\mu^{a}\nabla_{a}\phi+\half h^{ab}
\nabla_{a}\nabla_{b}\phi\right)dt + \sigma^{a}_{i}\nabla_{a}\phi
dW^{i}_{t} . 
\label{eq:7}
\end{eqnarray}
The probability law for $x_t$ is 
characterised by a density function $\rho(x,t)$ which satisfies 
the Fokker-Planck equation 
$\partial\rho/\partial t=-\nabla_{a}(\mu^{a}\rho)+
\half \nabla_{a}\nabla_{b}(h^{ab}\rho)$. 
Such a diffusion is nondegenerate if $h^{ab}$ is of 
maximal rank. In particular, if $g_{ab}$ is a Riemannian metric 
on ${\mathfrak M}$ and $\nabla_{a}$ is the associated Levi-Civita 
connection, then if $h^{ab}=\sigma^{2}g^{ab}$, the process $x_t$ 
is a Brownian motion with drift on ${\mathfrak M}$, with 
volatility parameter $\sigma$. If $h^{ab}$ is not of maximal 
rank, then the diffusion is degenerate, which is the 
case of relevance to the present consideration. 

Our intention is to generalise the Schr\"odinger dynamics to 
a stochastic process on a nonlinear quantum state manifold 
${\mathfrak M}$. Specifically, we consider the stochastic 
reduction model of Hughston \cite{hughston}, for which the 
dynamical trajectories are governed by the following stochastic 
differential equation: 
\begin{eqnarray} 
dx^{a}_{t}=\left( 2\omega^{ab}\nabla_{b}H-\quat 
\sigma^{2}\nabla^{a}V^{H}\right)dt+\sigma\nabla^{a}H dW_{t} . 
\label{eq:8}
\end{eqnarray}
When ${\mathfrak M}$ is the state space $\Gamma$ of linear 
quantum mechanics, then (\ref{eq:8}) has the following 
interpretation. The first term in the drift generates the 
unitary part of the evolution, while the second term creates 
a tendency for the system to evolve 
to a state of lower energy variance. The volatility term is given 
by the gradient of the Hamiltonian, and generates fluctuations 
that die down as the system approaches an eigenstate. The 
parameter $\sigma$ controls the magnitude of the fluctuations. 
Starting from any initial state, the state vector collapses to an 
energy eigenstate, with collapse probability given by the Dirac 
transition probability. Furthermore, if the evolution of the density 
function $\rho(x,t)$ associated with the process (\ref{eq:8}) is 
lifted to ${\cal H}$, we recover the Lindblad form of the density 
matrix dynamics \cite{gisin,adler2}. 

In the case of a general nonlinear quantum phase space 
${\mathfrak M}$, the stochastic process (\ref{eq:8}) can be carried 
over directly, and we obtain the following characterisation of the 
dynamics of the energy: 

{\bf Theorem 1}. {\sl The Hamiltonian process 
$H_{t}=H(x_{t})$ is a martingale, given by the stochastic 
integral $H_{t} = H_{0} + \sigma \int_{0}^{t} V_{s} dW_{s}$, 
where $V_{t}=V^{H}(x_{t})$.} 

Therefore, under the stochastic dynamics (\ref{eq:8}), the energy 
of the system is weakly conserved in the sense that the expectation 
of the value of the Hamiltonian at any future time is given by the 
initial value $H_{0}$. In particular, if reduction occurs, the 
martingale property ensures that the expectation of the terminal 
value of the energy, which is given by the sum of the energy 
eigenvalues weighted by the associated transition probabilities, is 
$H_{0}$. This in turn justifies the interpretation of $H(x)$ as the 
expectation of the energy in the given state $x\in{\mathfrak M}$. 
The proof of Theorem 1 follows from an application of Ito's formula 
(\ref{eq:7}). The fact that $H_{t}$ is a martingale does not imply 
that reduction occurs. For a reduction to energy eigenstates, we 
require $\lim_{t\rightarrow\infty}V_{t}=0$. To determine the 
circumstances under which this occurs, we consider the dynamics 
of $V_t$.  

{\bf Lemma 4}. {\sl The process $V_{t}=
\nabla_{a}H\nabla^{a}H$ satisfies } 
\begin{eqnarray} 
dV_{t} &=& \sigma^{2}(\nabla^{a}H\nabla^{b}H\nabla^{c}H
\nabla_{a}\nabla_{b}\nabla_{c}H)dt \nonumber \\ 
& &\ + \sigma(\nabla^{a}H\nabla_{a}V)dW_{t} .  
\label{eq:11}
\end{eqnarray}

The proof is as follows. We note that according to Ito's formula 
(\ref{eq:7}) we have 
\begin{eqnarray} 
dV_{t} &=& \left( 2\omega^{ab}\nabla_{a}V\nabla_{b}H - \quat 
\sigma^{2}\nabla_{a}V\nabla^{a}V \right. \nonumber \\ 
& & \hspace{-0.5cm} + \left. \half \sigma^{2}\nabla^{a}H\nabla^{b}H 
\nabla_{a}\nabla_{b}V \right)dt 
+ \sigma \nabla^{a}H\nabla_{a}VdW_{t} . 
\label{eq:12}
\end{eqnarray}
The first term in the drift vanishes by Lemma 3. The remaining 
two terms in the drift combine to yield (\ref{eq:11}), because 
$\nabla^{a}H\nabla^{b}H
\nabla_{a}\nabla_{b}V=2\nabla^{a}H\nabla^{b}H\nabla^{c}H
\nabla_{a}\nabla_{b}\nabla_{c}H+
\frac{1}{2}\nabla_{a}V\nabla^{a}V$. 

For state reduction, we need to show that the 
drift of $V_{t}$ is negative. To obtain a suitable criterion for 
this we proceed as 
follows. If $X_{a}$ is a Killing field, the cyclic identity 
$\nabla_{[a}\nabla_{b}X_{c]}=0$ implies that 
$\nabla_{c}\nabla_{a}X_{b}=R_{abc}^{\ \ \ d}X_{d}$, 
where the Riemann tensor is defined by 
$\nabla_{a}\nabla_{b}A_{c}-\nabla_{b}\nabla_{a}A_{c}
= - R_{abc}^{\ \ \ d}A_{d}$ for any vector field $A_{a}$. As 
shown in \cite{cirelli} we thus have: 

{\bf Lemma 5}. {\sl If $F(x)$ is an observable on 
${\mathfrak M}$, then  
$\nabla_{a}\nabla_{b}\nabla_{c}F = -J_{b}^{\ p}R_{apc}^{\ \ \ q}
J_{q}^{\ r}\nabla_{r}F$. } 

Next, we define the holomorphic sectional curvature ${\cal K}_{H}$ 
of the K\"ahler manifold ${\mathfrak M}$ with respect to the 
$J$-invariant plane $\nabla^{[a}HJ^{b]}_{\ c}\nabla^{c}H$ by the 
formula 
\begin{eqnarray} 
{\cal K}_{H} = - \frac{R_{apbq}J^{p}_{\ c}J^{q}_{\ d}
\nabla^{a}H\nabla^{b}H\nabla^{c}H\nabla^{d}H}
{(\nabla_{a}H\nabla^{a}H)^{2}} .
\label{eq:14}
\end{eqnarray}
The meaning of ${\cal K}_{H}$ is as follows. At each point 
$x\in{\mathfrak M}$ we consider the tangent plane spanned by 
vectors $\nabla^{a}H$ and $J^{b}_{\ c}\nabla^{c}H$. 
Then the totality of the geodesic curves tangent to this plane 
at $x$ forms a two dimensional surface in ${\mathfrak M}$, and 
the Gauss curvature of this surface at $x$ is ${\cal K}_{H}$.  

{\bf Theorem 2}. {\sl If ${\cal K}_{H}>0$, then 
$V^{H}$ is a supermartingale and {\rm (\ref{eq:8})} 
is a reduction process.} 

The proof is by virtue of Lemmas 4 and 5. Writing ${\cal K}_{t}
={\cal K}_{H}(x_t)$, we deduce that 
\begin{eqnarray}
V_{t}=V_{0}-\sigma^{2}\int_{0}^{t}{\cal K}_{s}V_{s}^{2}ds 
+\sigma\int_{0}^{t}\nabla_{a}H\nabla^{a}V dW_{s} , 
\label{eq:15}
\end{eqnarray}
and thus ${\mathbb E}[V_{t}|{\cal F}_{s}]\leq V_{s}$, the 
supermartingale condition. In particular, if we write 
${\bar V}_{t}={\mathbb E}[V_{t}]$ then it follows from 
(\ref{eq:15}) that $d{\bar V}_{t}/dt=-\kappa\sigma^{2}
{\bar V}_{t}^{2}(1+\eta_{t})$, where ${\bar V}_{t}^{2}\eta_{t}=
{\mathbb E}[(V_{t}-{\bar V}_{t})^{2}]+\kappa^{-1}
{\mathbb E}[({\cal K}_{t}-\kappa)V_{t}^{2}]$, and $\kappa=
\inf_{{\mathfrak M}}{\cal K}_{H}$. Integrating, we obtain 
${\bar V}_{t}=V_{0}/(1+\kappa\sigma^{2}V_{0}(t+\xi_{t}))$, 
where $\xi_{t}=\int_{0}^{t}\eta_{s}ds$. The Hamiltonian 
sectional curvature is positive iff $\kappa>0$, in which case 
$\xi_{t}\geq0$. It follows that 
${\bar V}_{t}\leq V_{0}/(1+\kappa\sigma^{2}V_{0}t)$, 
and thus $\lim_{t\rightarrow\infty}V_{t}=0$ almost surely. 
Therefore, wave function collapse on 
the nonlinear state space is guaranteed if the holomorphic 
sectional curvature is positive. The characteristic 
reduction time-scale is $\tau=(\kappa\sigma^{2}V_{0})^{-1}$, 
and for $t\gg\tau$ the uncertainty is reduced to a fraction of 
its initial value. 

Now we are in a position to determine the relationship between the 
initial energy uncertainty $V_{0}=V^{H}(x_{0})$ and the terminal 
variance of the energy as a result of a reduction. It follows 
from Theorem 1, together with the Ito isometry, that ${\mathbb E}
\left[(H_{t}-H_{0})^{2}\right] = \sigma^{2}{\bar Q}_t$, where 
$H_0={\mathbb E}[H_t]$, ${\bar Q}_t={\mathbb E}[Q_t]$, and 
$Q_{t}=\int_{0}^{t}V_{s}^{2}ds$. On the other hand, if $\kappa>0$ 
then from (\ref{eq:15}) we obtain ${\bar V}_{t}\leq V_{0}-
\kappa\sigma^{2}{\bar Q}_t$. Furthermore, if 
$\lambda=\sup_{{\mathfrak M}}{\cal K}_{H}$ and $\lambda>0$, then 
(\ref{eq:15}) implies that ${\bar V}_{t}\geq V_{0}-\lambda
\sigma^{2}{\bar Q}_{t}$. 
Therefore, by Theorem 2, if ${\cal K}_{H}>0$, we obtain the 
following bounds for the terminal energy variance:  
\begin{eqnarray}
\frac{V_{0}}{\kappa} \geq \lim_{t\rightarrow\infty} 
{\mathbb E}\left[(H_{t}-H_{0})^{2}\right] \geq \frac{V_{0}}{\lambda} . 
\label{eq:21}
\end{eqnarray}
In particular, if ${\mathfrak M}=\Gamma$, it follows from the 
relation 
$R_{abcd}=-\frac{1}{4}(g_{ac}g_{bd}-g_{bc}g_{ad}+\omega_{ac}
\omega_{bd}-\omega_{bc}\omega_{ad}+2\omega_{ab}\omega_{cd})$ that 
${\cal K}_{H}=1$ and $V_{0}$ is the terminal energy dispersion. 

An important issue in the consideration of state reduction 
processes is whether an energy-based dynamics suffices. To address 
this issue we examine the processes induced by (\ref{eq:8}) for 
observables other than $H$. 

{\bf Theorem 3}. {\sl If an observable $F$ commutes with the 
Hamiltonian, then under the stochastic Schr\"odinger dynamics 
{\rm (\ref{eq:8})} the process $F_{t}=F(x_{t})$ is a martingale.} 

The proof follows as a consequence of Ito's lemma with an 
application of Lemmas 2 and 3. Theorem 3 generalises a result 
of \cite{adler} obtained in the case ${\mathfrak M}=\Gamma$. To 
determine whether the system necessarily collapses to an 
eigenstate of $F$ under (\ref{eq:8}) we require the concept of 
holomorphic bisectional curvature \cite{koba}. 

{\bf Theorem 4}. {\sl If the observable $F$ commutes with the 
Hamiltonian, then the stochastic equation for $V^{F}_{t}$ is  
\begin{eqnarray} 
dV^{F}_{t}=-\sigma^{2} {\cal K}_{FH}V^{F}V^{H} dt 
+\sigma\nabla_{a}H\nabla^{a}V^{F} dW_{t} , 
\label{eq:17}
\end{eqnarray}
where the holomorphic bisectional curvature of ${\mathfrak M}$ with 
respect to the $J$-invariant planes determined by $F$ and $H$ is 
defined by 
\begin{eqnarray} 
{\cal K}_{FH} = - 
\frac{R_{apbq}J^{p}_{\ c}J^{q}_{\ d}
\nabla^{a}F\nabla^{b}H\nabla^{c}F\nabla^{d}H}
{\nabla_{a}F\nabla^{a}F\nabla_{b}H\nabla^{b}H} . 
\label{eq:18}
\end{eqnarray}
} 

\vspace{-0.2cm}
To prove this result, we use Ito's formula (\ref{eq:7}) to obtain 
\begin{eqnarray}
dV^{F}_{t} &=& \half \sigma^{2} \left( \nabla^{a}H\nabla^{b}H
\nabla_{a}\nabla_{b}V^{F} \right. \nonumber \\ 
& & \hspace{-0.4cm} -\left. 
\nabla^{a}\nabla^{b}H\nabla_{a}H\nabla_{b}V^{F} 
\right) dt + \sigma\nabla^{a}H\nabla_{a}V^{F} dW_{t} . 
\label{eq:19} 
\end{eqnarray} 
Then, by use of Lemmas 2, 3, 5, a calculation shows that the two 
terms in the drift combine to give (\ref{eq:17}). As a consequence 
we have: 

{\bf Theorem 5}. {\sl If the holomorphic bisectional curvature is 
positive, then for any observable $F$ commuting with the 
Hamiltonian, the associated dispersion $V^{F}_{t}$ is a 
supermartingale, and {\rm (\ref{eq:8})} is a reduction process 
for $F$.}  

In the linear theory, this result is 
intuitively expected, because the eigenstates of $H$ also 
diagonalise any commuting observable $F$. Indeed, when 
${\mathfrak M}=\Gamma$ we have 
${\cal K}_{FH}=\frac{1}{2}(1+\cos^{2}\theta)$, where $\theta$ is 
the angle between the vectors $\nabla^{a}H$ and $\nabla^{a}F$, 
from which it follows that $\frac{1}{2}<{\cal K}_{FH}\leq1$, and 
thus reduction is guaranteed.

DCB gratefully acknowledges financial support from The Royal 
Society. 

$*$ Electronic mail: dorje@ic.ac.uk 
 
$\dagger$ Electronic mail: lane.hughston@kcl.ac.uk

\end{document}